\documentstyle[12pt]{article}
\newcommand{\tr}{\mbox{Tr}}
\newcommand{\be}{\begin{eqnarray}}
\newcommand{\ee}{\end{eqnarray}}
\newcommand{\beq}{\begin{equation}}
\newcommand{\eeq}{\end{equation}}

\newcommand{\fett}[1]{ \bf {#1} \rm }
\newcommand{\ol}{\overline}
\newcommand{\ep}{\varepsilon}
\newcommand{\C}{C\hspace{-0.7em}I\:}
\newcommand{\R}{R\hspace{-1em}I\ \ }
\newcommand{\I}{I\hspace{-0.4em}I\:}
\newcommand{\1}{1\hspace{-0.3em}\rm I \:}
\newcommand{\M}{M\hspace{-1.2em}I\ \ }

\newcommand{\Sp}{S\hspace{-0.6em}I\ }
\newcommand{\T}{T\hspace{-0.8em}I\ }
\newcommand{\sM}{M\hspace{-0.9em}I\ }
\newcommand{\sR}{R\hspace{-0.7em}I\ }

\begin{document}

\thispagestyle{empty}
\hfill MPI-PhT/96-1 (March 96)

\begin{center}

{\Huge Operational Discrete Symmetries and \\ CP-Violation  \\}
{\large
\vspace{3ex} 
 M. Haft, H. Saller\footnote{
e-mail adresses: \\
M.H.: mah@mppmu.mpg.de \\
H.S.: hns@mppmu.mpg.de 
} \\[5ex]
{\it Max-Planck-Institut f\"{u}r Physik \\
 (Werner-Heisenberg-Institut) \\
 80805 Munich (Germany) }
}

\end{center}

\vspace{5ex}
\centerline{\bf Abstract}

The discrete symmetries of the Lorentz group are on the one hand a `complex' 
interplay between linear
and anti-linear operations on spinor fields and on the other hand 
simple linear reflections of the Minkowski space. We define 
operations for $T$, $CP$ and $CPT$ leading to both kinds of actions. 
These operations extend the action of $SL(2,\C)$, representing 
the action of the proper orthochronous Lorentz group $SO^+(1,3)$ 
on the Weyl spinors, to an action of the full group
$O(1,3)$. But it is more instructive to reverse the arguments. The action of
$O(1,3)$ is the natural way how $SL(2,\C)$ together with its conjugation 
structure acts on Minkowski space. 

Focusing on the symmetries of these (anti-)linear
operations we can for example distinguish between $CP$-invariant and 
$CP$-violating symmetries. 
This is important if gauge symmetries are included. It turns out that, 
contrary to the general belief, $CP$
and $T$ are not compatible with $SU(n)$ for $n \geq 3$, especially with
$SU(3)_{\mbox{colour}}$ or with the $U(3)$-Cabibbo-Kobayashi-Maskawa matrix. 

%\end{abstract}

%\thispagestyle{empty}

\newpage
\pagenumbering{arabic}

The history of the discrete symmetries was a history of surprises. For example
when C.S. Wu discovered parity violation (after theoretical advice 
given by Lee and Yang), Wolfgang Pauli wrote to his former assistant Viktor
Weisskopf: "What shocks me is not the fact that `God is just left-handed' but
the fact that in spite of this He exhibits Himself as left/right symmetric when
He expresses Himself strongly. In short, the real problem now is why the strong
interactions are left/right symmetric. How can the strength of an interaction
produce or create symmetry groups, invariances or conservation laws? This
question prompted me to my premature and wrong prognosis. I don't know any good
answer to that question but one should consider that already there exists a
precedent: the rotational group in isotopic spin-space, which is not valid for
the electromagnetic field. One does not understand either why it is valid at
all. It seems that there is a certain analogy here!" \cite{Kro 64}.
Even more unexpected was the discovery of $CP$-violation by Finch et al.

Why was there such a surprise? Beginning with
the discovery of spin by Stern and Gerlach and with the theoretical work of
Dirac and Weyl the `real version' of the Lorentz group, i.e. $O(1,3)$, lost more
and more of its fundamental meaning and should have been replaced by  
$SL(2,\C)$. But therein parity is not defined. Only after the discovery
of parity violation the Weyl theory became familiar. Nowadays the Standard
Model is written in Weyl spinors. Is there a similar hint for $CP$-violation?
In order to answer this question one has to take a look at the representation 
structure of the
Lorentz group on the Weyl spinors. Is it possible to understand the discrete
part of the Lorentz group like the continuous part? In the first three sections
we define the discrete symmetries as (anti-)linear operations within the
different kinds of Weyl spinors. Their action on the Cartan (bispinor)
representation of the Minkowski space is the familiar action of the
discrete symmetries on Minkowski space. Next we show that these operations 
lead also to the discrete symmetry operations on Dirac spinor fields. 

The discrete symmetry operations on the Weyl spinors are deeply connected with
the spin and boost structure of $SL(2,\C)$. Including also inner 
symmetries we show that the discrete symmetries are not 
compatible with every representation of the inner symmetry groups. 
This holds especially for
$CP$ and $T$ in an $SU(3)$ gauge theory. It is interesting that for the same
reason the $U(3)$-Cabibbo-Kobayashi-Maskawa matrix breaks $CP$ invariance. 

\section{The Discrete Factors \\ of the Lorentz Group $O(1,3)$} \label{KO13}
The Minkowski time-space translations $\M$ form a 4-dimensional real vector 
space with  
bilinear form of signature (1,3). This bilinear form $\eta$ 
is usually called the Lorentz metric. It is left invariant by the 
action of 
the Lorentz group $O(1,3)$. The Lorentz group is not simply connected. This is 
expressed in its semidirect and direct product structure of the sign group
$\I_2 \cong \{\pm \1 \}$ with the proper orthochronous Lorentz group $SO^+(1,3)$:
\beq
	O(1,3)\ \cong\ \I_2^{T} \times_s \left( \I_2^{CPT} \times 
		SO^+(1,3) \right). \label{O13}
\eeq
Here $\times$ denotes the direct product and $\times_s$ the semidirect 
product\footnote{Contrary to the usual 
mathematical notation the normal subgroup is written as second 
factor in the semidirect product. In this notation the Poincar\'e group
reads $O(1,3) \times_s \R^4$. This notation reflects the action of the first
factor group onto the normal subgroup as second factor, whereas there is 
no reverse action.}. 

The
discrete factors are labelled according to their representation on the Minkowski space
\beq
	\I_2^{CPT} \ \cong \ \{ \pm \1_4 \} ,  \eeq \beq
	\I_2^T \ \cong \ \{ \1_4, \left( \begin{array}{cc}
						-1 & \\
						& \1_3
					  \end{array} \right) \}  \label{I2a}.
\eeq
The element $-\1_4$ of the representation of $\I_2^{CPT}$ is called
the strong reflection and is the representation of CPT on the Minkowski
space. The factor $\I_2^T$ contains the operation of the
time reversal T. However, there is no canonical way
to decompose the Minkowski space $\M$ into time $\T$ and space $\Sp$, i.e. 
$\T \oplus \Sp$; this is like the Sylvester form of the Lorentz metric, 
$\eta = 
{\scriptsize \left( \begin{array}{cc} 1 & \\ & -\1_3 \end{array} \right) } $, 
which is 
only one possible form and in no way distingushed if there is no rest system. 
But without such a decomposition, there is no representation of
$\I_2^T$ possible like in (\ref{I2a}). When given one rest system, then
in a different (boosted) system, the space-time decomposition is different
and therefore the representation of $\I_2^T$ is different. 
This is due to the semidirect
product structure, since $\I_2^T$ commutes with the rotations $SO(3) 
\subset
SO^+(1,3)$, but not with the boosts $SO^+(1,3)/SO(3)$. 

The symmetry group of the Euclidean space is $O(3)$, the direct product group 
of rotations and space reflection (parity P)
\beq
	O(3) \ = \ \I_2^P \times SO(3), \ \ \I_2^P \cong \{ \pm \1_3 \}.
\eeq
The group $O(3)$ can be embedded in the Lorentz group in two different ways
\be
	O(3) \ = \ \I_2^P \times SO(3)  
	& \subset &  \I_2^{CPT} \times SO^+(1,3)\ = \ SO(1,3) \\
	O(3) \ = \ \I_2^P \times SO(3)
	& \subset &  \I_2^{CP} \times_s SO^+(1,3)\ \cong \ O^+(1,3) 
\ee
with $\I_2^{CP} \subset \I_2^T \times_s \I_2^{CPT}$, $\I_2^{CP} \cong \{ \1_4, 
{\scriptsize \left( \begin{array}{cc} 1 & \\ & -\1_3 \end{array} \right) } \}$.
Thus the embedding of parity is not unique. As far as no embedding is
distinguished one should not identify the operation $
{\scriptsize \left( \begin{array}{cc} 1 & \\ & -\1_3 \end{array} \right) }$ 
with the representation of $P$.  
What we will show in the following is that this operation 
can be identified with CP.  We will also show which
structure, additional to the Lorentz group, is needed to define CP and T and
how this is related to a space-time decomposition.  

\section{Spinor Representation of Minkowski Space}
The proper orthochronous Lorentz group $SO^+(1,3)$ is the $D^{(\frac{1}{2}|
\frac{1}{2})}$
representation of its covering group\footnote{In the following the group 
$SL(2,\C)$ is always regarded as a 6-dimensional real Lie group: $SL(2,\C)
=  SL(2,\C)_{\sR}$. As 6-dimensional real Lie group we refer to it also as
Lorentz group.} $SL(2,\C)$, a tensor product  
representation of the two fundamental $SL(2,\C)$-representations. 
What follows in this section is
the basis-independent definition of this representation and its representation 
space -- 
the Cartan (or bispinor) representation of the Minkowski space \cite{Car}. 

The reader familiar with basis-independent complex 
representations of real Lie-groups can read this section 
only for notations.

In general a complex vector space appears in a fourfold way. Every vector
space $V$ has its dual space $V^T$, the linear forms on $V$. In addition 
to each 
complex vector space with action of the field $\C$ (scalar multiplication)
\beq
	z \bullet v \ = \ z v \ \ z \in \C, \ v \in V ,
\eeq
there is a complex conjugate or anti-space\footnote{The complex anti-space is a
special case of a modul over a ring, where the ring has a canonical
automorphism structure (compare \cite{Bou 89II}). The anti-space is also
introduced in some physical literature, for example in \cite{BT 88} and in 
the appendix (2nd edition and later) of \cite{SU 92}. In the case of
representation theory of the Lorentz group it is the vector space of the
vectors with, due to Weyl, dotted indices.}  
$\ol{V}$ with complex conjugate action
\beq
	z \bullet v \ = \ \bar{z} v \ \ z \in \C, \ v \in \ol{V} .
\eeq
The two vector spaces $V$ and $\ol{V}$ are identical when regarded as 
additive groups, 
they have to be distinguished when regarded as vector spaces. 

All together we have a quartet of complex vector spaces, $V, V^T, \ol{V}$ and 
$\ol{V}^T$, where the anti-spaces and spaces are related to each other by the 
complex conjugation. \\
\linebreak
\setlength{\unitlength}{0.6cm}
\begin{picture}(10,6.5)(-4,0)
        \put(0.5,0.5){\makebox{$\ol{V}$}}
        \put(0.5,5.0){\makebox{$V$}}
        \put(5.5,5.0){\makebox{$V^T$}}
        \put(5.5,0.5){\makebox{$\ol{V}^T$}}
        \put(1.0,5.1){\vector(1,0){4}}
        \put(1.0,0.8){\vector(1,0){4}}
        \put(0.6,4.5){\vector(0,-1){3}}
        \put(5.6,4.5){\vector(0,-1){3}}
        \put(2.5,5.4){\makebox{$\xi$}}
        \put(2.5,0.0){\makebox{$\bar{\xi}$}}
        \put(-0.5,3.0){\makebox{$co_V$}}
        \put(6.1,3.0){\makebox{$co_{V^T}$}}
\end{picture}
\linebreak 
The complex conjugation $co_V$ acts on the additive groups 
underlying $V$ and $\ol{V}$ 
as the identity, but on the vector spaces it has the anti-linear property 
\be
	co_V (\alpha v) \ = \ \bar\alpha co_V(v), \ \ v \in V, \alpha \in \C  .
\ee
We refer to it as the canonical conjugation of the vector spaces. 

In addition, there may exist an isomorphism $\xi$ between the
dual vector spaces $V$ and $V^T$. The corresponding isomorphism $\bar{\xi}$ 
between $\ol{V}$ and $\ol{V}^T$ is given by the canonical conjugation
\be
	\bar{\xi} \  = \  co_{V^T} \xi co_V^{-1}. 
\ee

There is an analogue fourfold structure 
of in general inequivalent representations of a group on these 
vector spaces. The group $GL(n,\C)$, $n > 1$, regarded as a real
$2n^2$-dimensional Lie-group, 
has four complex $n$-dimensional fundamental
representations. With the defining representation 
$D_V(g) = D(g) = g \in GL(n,\C)$ 
given on $V \cong \C^n$ one has the three partners: 
\be
	D_{V^T}(g) &=& D(g)^{-1T} \ =: \ \check{D}(g)  \label{DVT}\\
	D_{\ol{V}}(g) &=& co_V D(g) co_V^{-1} \ =: \ \bar{D}(g)  \\
	D_{\ol{V}^T}(g) &=& \bar{D}(g)^{-1T} \ =: \ \check{\bar{D}}(g) 
		\label{DVTQ}.   
\ee

In the case of $s \in SL(2,\C)$, the dual representations on $V$ and $V^T$
or $\ol{V}$ and $\ol{V}^T$ are equivalent 
with the volume form\footnote{In 
general the $SL(n,\C)$-invariant volume form is multi-linear and totally 
antisymmetric. Only in the 
special case $n = 2$ it is a bilinear form and therewith it is equivalent to a 
dual isomorphism.} or spinor metric, 
\[
	\ep : V \longrightarrow V^T , \] \beq
	\ep(e^A) \ = \ \ep^{AB} \check{e}_B   \label{ep} ,
\eeq 
$e^A$ and $\check{e}_A$ being dual bases of $V$ and $V^T$, resp., and
$\ep^{AB} = {\scriptsize \left( \begin{array}{cc} 0 &  1 \\ -1 & 0 
\end{array} \right) }$ the matrix representation of the dual
isomorphism\footnote{It is sometimes denoted as $i \sigma_2$.} in
this basis. The equivalence of the $SL(2,\C)$ representations is expressed by  
\be
	\check{D}(s) &=& \ep D(s) \ep^{-1}  \\
	\check{\bar{D}}(s) &=& \bar\ep \bar{D}(s) \bar\ep^{-1}  .
\ee
The two fundamental representations are chosen as 
\be
	D(s) &=:& D^{(\frac{1}{2}|0)}(s)  \\
	\check{\bar{D}}(s) &=:& D^{(0|\frac{1}{2})}(s) ,
\ee
which are the left-handed and right-handed Weyl representations, resp. 
They are connected by the action of the canonical conjugation together with the
transposition $(\cdot)^\times := \bar{(\cdot)}^T$. With equations (\ref
{DVT}) to (\ref{DVTQ}) this gives  
\beq
	D(s)^\times \ = \ \check{\bar{D}}(s)^{-1} \label{U22cc}.
\eeq
Therefore we refer to this operation as canonical conjugation on the
representations. 

The Dirac vector space $V_D := V \oplus \ol{V}^T \cong \C^4$ contains the two  
fundamental
Weyl vector spaces. Therewith the Dirac representation of the Lorentz group
is the direct sum of both fundamental Weyl representations:
 $D_D(s) = D(s) \oplus \check{\bar{D}}(s)$. 
This representation lies in the complex 16-dimensional 
endomorphism algebra of the Dirac vector space, called the Dirac 
endomorphisms\footnote{A subalgebra of the Dirac endomorphisms is the real
16-dimensional Dirac algebra, symmetric with respect to the canonical
conjugation.} 
\beq
	\mbox{end}(V_D) \ \cong \ V_D \otimes V_D^T \ \cong \ 
	\left(
		\begin{array}{cc}
		V \otimes V^T & V \otimes \ol{V}  \\
		\ol{V}^T \otimes V^T & \ol{V}^T \otimes \ol{V} 
		\end{array}
	\right)
\eeq
The canonical conjugation of the underlying Weyl quartet gives a
conjugate linear reflection 
of the Dirac endomorphisms. Its invertible elements with the property 
$f^\times = f^{-1}$
are elements of the indefinite unitary group $U(2,2)$. Remembering eq.(\ref{U22cc}) 
one can see, that the Dirac representation is an embedding of the
Lorentz group into $U(2,2)$\footnote{Via the Dirac construction 
of a doubled vector space the whole $GL(n,\C)$ is embedded as
an indefinite unitary reducible representation in the group $U(n,n)$. 
In other words,
this gives an indefinite unitary representation of $GL(n,\C)$ \cite{Sall}.}. 
The quotient group $U(2,2)/U(1) \cong SO(2,4)$
 - the conformal group - contains 
the whole Poincar\'e group. Therefore in the Dirac endomorphisms there is a
vector subspace with the properties of the Minkowski translations being
anti-symmetric with respect to the canonical conjugation. We identify 
\beq
	\M \ = \ \left\{ x \in V \otimes \ol{V} | x^\times = -x \right\} 
\eeq
as the Cartan representation of the Minkowski space. This is a real 
subspace of the
linear mappings from $\ol{V}^T$ (right-handed Weyl spinors) to $V$ (left-handed
Weyl spinors) $x: \ol{V}^T \rightarrow V$. 

For these $2 \times 2$ dimensional mappings we can choose an appropriate  
basis ${\bf e}^\mu, \mu = 0, .. , 3$ with 
$({\bf e}^\mu)^\times = - {\bf e}^\mu$. 
One possible matrix representation is given with the Pauli matrices by 
$i \rho^\mu := (i \1, \it i \sigma^i)$, referred to as Weyl basis. 
In this basis an element of the Minkowski space
\beq
	\M \ = \ \left\{ x_\mu \ {\bf e}^\mu | x_\mu \in \R
		\right\}  \label{MB1}
\eeq
has the matrix representation 
\beq
	x \ \cong \ x_\mu i \rho^{\mu}\ = \
	i \ \left(
	\begin{array}{cc}
		x_0 + x_3 & x_1 - ix_2  \\
		x_1 + i x_2 & x_0 - x_3
	\end{array} \right) \label{MB2}
\eeq
which was first introduced by E.Cartan \cite{Car} (further
developments are in \cite{Wae 32,BW 35}). This matrix representation
should always be regarded as embedded in the Dirac endomorphisms
\[
	x_\mu \ {\bf e}^\mu\  \cong\   \left(
		\begin{array}{cc}
		0 &  x_\mu i \rho^{\mu} \\
		0 & 0 
		\end{array}
	\right) .
\]
Hence considered as Dirac endomorphisms the Minkowski space has a nilpotent
product.

We have to emphasize the
basis dependence of the definition of the Minkowski space in the 
representation (\ref{MB2}). The Weyl basis is already a basis
in which the Lorentz metric has the form $\eta = 
{\scriptsize \left( \begin{array}{cc} 1 & \\ & -\1_3 \end{array} \right) } $. 
(see app. \ref{AA}.)
This basis anticipates a space-time
decomposition and is appropriate to define the $CP$ and $T$ operations. The 
space-time
decomposition without an anticipating basis is given in sect. \ref{EK}. 
	
To obtain the action of $SL(2,\C)$ on the Cartan representation of 
Minkowski space we use a method called induced action\footnote{This induced
action is more general than the method of induced representations given by Mackey
\cite{Mac}. The Mackey theory can be formulated in this
language.}: In general when there are two $G$-sets $S_1$ and $S_2$, 
defined as two sets 
with an action $\rho_1(g)$ and $\rho_2(g)$ of a group $g \in G$ resp., 
$S_1, S_2 \in \mbox{set}_G$ \cite{Bou 89I}, then there is an induced
$G$-action on the mappings $S_2^{S_1}$ between these two sets, 
$f : S_1 \longrightarrow S_2$, 
i.e. $S_2^{S_1} \in \mbox{set}_G$. The induced action can
be characterized by the commutative diagram
\linebreak
\setlength{\unitlength}{0.6cm}
\begin{picture}(10,6.5)(-4,0)
        \put(0.5,0.5){\makebox{$S_2$}}
        \put(0.5,5.0){\makebox{$S_1$}}
        \put(5.5,5.0){\makebox{$S_1$}}
        \put(5.5,0.5){\makebox{$S_2$}}
        \put(1.0,5.1){\vector(1,0){4}}
        \put(1.0,0.8){\vector(1,0){4}}
        \put(0.6,4.5){\vector(0,-1){3}}
        \put(5.6,4.5){\vector(0,-1){3}}
        \put(2.5,5.4){\makebox{$\rho_1(g)$}}
        \put(2.5,0.2){\makebox{$\rho_2(g)$}}
        \put(-0.3,3.0){\makebox{$f$}}
        \put(6.1,3.0){\makebox{$g \bullet f = f_g$}}
\end{picture}
\linebreak   
where $g \bullet f$ denotes the action of $g \in G$ on the mapping $f$: 
\beq
	g \bullet f \ = \ f_g \ = \ \rho_2(g) \circ f \circ \rho_1(g)^{-1}  
		\label{fg}.
\eeq
This inducing construction is necessary to obtain the action of the discrete
symmetry operations on the Minkowski space in the next section. It can also
be used for the action of $SL(2,\C)$ on the Cartan representation of the
Minkowski space: 
Substitute the sets $S_i$ by the left- and right-handed Weyl vector
spaces, $V$ and $\ol{V}^T$, resp., the actions $\rho_i(g)$ by the
representations $D(s)$ and $\check{\bar{D}}(s)$ and the mappings $f$ by
elements of the Minkowski space $x$. Then the action of $SL(2,\C)$ on Minkowski
space is given with eq.(\ref{fg}) by
\beq
	s \bullet x \ =\ D^{(\frac{1}{2}|0)}(s) \circ x \circ 
		D^{(0|\frac{1}{2})}(s)^{-1} \ =\
		D^{(\frac{1}{2}|\frac{1}{2})}(s).x  \ \ . \label{LorentzDarst}
\eeq
Because of the isomorphism 
\beq
	SO^+(1,3) \ \cong \ SL(2,\C)/\I_2  ,
\eeq
$\I_2 \cong \{ \pm \1_{\rm 2} \}$, the above representation is a faithful 
representation of $SO^+(1,3)$ only. In the Weyl basis of the Minkowski space
this representation leads to the familiar Lorentz matrices 
$\Lambda(s)^\mu_\nu$ as matrix representation
\beq
	D^{(\frac{1}{2}|\frac{1}{2})}(s).x \ \cong \ 
	x_\mu \Lambda(s)^\mu_\nu i \rho^\nu .
\eeq
	
\section{The Full Lorentz Group from Actions on the Weyl Spinors}\label{S3}
The actions of $SL(2,\C)$ can be regarded as the action of $SO^+(1,3)$ on the
Weyl spinor spaces. 
Or vice versa, $SO^+(1,3)$ is the natural action of 
$SL(2,\C)$ on the Cartan representation of the Minkowski space. This picture
seems to fail 
for the discrete parts of the full group $O(1,3)$. However, there are additional
operations within the complex quartet of the Weyl vector spaces (both 
linear and anti-linear) acting on the Cartan representation of the Minkowski
space as (linear) automorphisms. 

According to Wigner the operations of $P$ and $C$ in quantum field
theory are linear and so is $CP$.
On the other hand $T$ and therewith $CPT$ are anti-linear operations
\cite{Wig 31, Wig 32}. 

One anti-linear action within the Weyl quartet reversing the Minkowski space 
is the canonical conjugation. With the property of the 
$SL(2,\C)$ representation acting conjugation compatible on the Minkowski space,
calculated from equations 
(\ref{U22cc}) and (\ref{LorentzDarst}), 
\beq
	\left( D^{(\frac{1}{2}|\frac{1}{2})}(s).x \right)^{\times} \ = \ 
		D^{(\frac{1}{2}|\frac{1}{2})}(s).x^\times   \label{Dx},
\eeq
i.e. the action of the canonical conjugation `commuting' with the action of the
Lorentz group
\[
	co (s \bullet x) \ = \ s \bullet co (x)  ,
\]
the `product group' generated by $co$ and $SL(2,\C)$ acts on the Minkowski 
space as the 
direct product group $\I_2^{CPT}
\times SO^+(1,3) \cong SO(1,3)$. In this context the canonical conjugation can
be regarded as the action of $CPT$ on the spinor spaces. In sect.\ref{WS} we
will show, that this action is also the $CPT$ action on spinor fields in
quantum field theory. 

According to the remarks at the end of sect.\ref{KO13} we need for the
representation of $T$ in addition to the
$SL(2,\C)$ compatible structures $co$ and $\ep$ a structural element being 
invariant
with the $SU(2)$ subgroup of $SL(2,\C)$, but not with the boosts. This new
structure is the anti-linear euclidian conjugation $\delta$, which in general 
is the invariant dual isomorphism of the positive unitary group $U(n)$. 
\[
	\delta : V \longrightarrow V^T, \ \ V \cong \C^2
\] \beq
	\delta(a_A e^A) \ = \ \bar{a}_A \delta^{AB} \check{e}_B \label{delta}
\eeq \beq
	\check{D}(u) \ = \ \delta D(u) \delta^{-1} , \ \ u \in U(n) \label{Un}
\eeq
or 
\beq
	D^\star(u) \ := \ \delta^{-1} D(u)^T \delta \ = \ D(u)^{-1} .
\eeq
Within the Weyl quartet the totality of all $SU(2)$ compatible operations is
characterized by the following diagram: \\
\linebreak
\setlength{\unitlength}{0.6cm}
\begin{picture}(10,6.5)(-4,0)
        \put(0.5,0.5){\makebox{$\ol{V}$}}
        \put(0.5,5.0){\makebox{$V$}}
        \put(5.5,5.0){\makebox{$V^T$}}
        \put(5.5,0.5){\makebox{$\ol{V}^T$}}
        \put(1.0,5.1){\vector(1,0){4}}
        \put(1.0,0.8){\vector(1,0){4}}
        \put(0.6,4.5){\vector(0,-1){3}}
        \put(5.6,4.5){\vector(0,-1){3}}
        \put(2.5,5.4){\makebox{$\ep , \delta$}}
        \put(2.5,0.0){\makebox{$\bar{\ep} , \bar{\delta}$}}
        \put(-0.5,3.0){\makebox{$co_V$}}
        \put(6.1,3.0){\makebox{$co_{V^T}$}}
\end{picture}
\linebreak 

Together with the volume form and the euclidian conjugation we can construct an 
anti-linear automorphism on
the Weyl vector space, $\delta^{-1} \circ \ep: V \longrightarrow V$, 
compatible
with $SU(2)$. The corresponding actions on the other partners of the quartet 
is given by
$\ep \delta^{-1}, \bar{\delta}^{-1} \bar\ep$ and $\bar\ep \bar\delta^{-1}$. 

With the concept of induced action we construct the action of these anti-linear
operations on the Cartan representation of Minkowski space   
\linebreak
\setlength{\unitlength}{0.6cm}
\begin{picture}(10,6.5)(-4,0)
        \put(0.5,0.5){\makebox{$V$}}
        \put(0.5,5.0){\makebox{$\ol{V}^T$}}
        \put(5.5,5.0){\makebox{$\ol{V}^T$}}
        \put(5.5,0.5){\makebox{$V$}}
        \put(1.0,5.1){\vector(1,0){4}}
        \put(1.0,0.8){\vector(1,0){4}}
        \put(0.6,4.5){\vector(0,-1){3}}
        \put(5.6,4.5){\vector(0,-1){3}}
        \put(2.5,5.4){\makebox{$\bar{\ep} \bar{\delta}^{-1} $}}
        \put(2.5,0.2){\makebox{$\delta^{-1} \ep$}}
        \put(-0.3,3.0){\makebox{$x$}}
        \put(6.1,3.0){\makebox{$(\delta^{-1} \ep) \bullet x = 
        	(\delta^{-1} \ep) \circ x \circ 
		(\bar{\ep} \bar{\delta}^{-1})^{-1}$}}
\end{picture}
\linebreak
This abstract operation can be concretized in the matrix representations of
equations (\ref{ep}), (\ref{MB2}) and (\ref{delta})  
\be
	( \delta^{-1} \ep ) \bullet x
		& \cong & \left( \begin{array}{cc}
					0 & 1 \\
					-1 & 0 
				\end{array} \right)
			\left[ i
			\left( \begin{array}{cc}
					x_0 + x_3 & x_1 - i x_2 \\
					x_1 + i x_2 & x_0 - x_3 
				\end{array} \right) \right]^*
			\left( \begin{array}{cc}
					0 & -1 \\
					1 & 0 
				\end{array} \right)  \nonumber  \\
		& \cong & \left( \begin{array}{cc}
					-x_0 + x_3 & x_1 - i x_2 \\
					x_1 + i x_2 & -x_0 - x_3 
				\end{array} \right) .
\ee
Hence the action of the combination of the volume form and the euclidian
conjugation implements the time reversal 
\beq
	( \delta^{-1} \ep ) \bullet x \ = \ x_\mu\ \Lambda(T)^\mu_\nu 
		{\bf e}^\nu \ \cong \ (-x_0, \vec{x}) , 
\eeq
with $\Lambda(T)^\mu_\nu = 
{\scriptsize \left( \begin{array}{cc} -1 & \\ & \1_3 \end{array} \right) } $.

Since the operation $\delta^{-1} \ep$ 
commutes with $SU(2)$ in $SL(2,\C)$, but not with the boosts $SL(2,\C)/SU(2)$,  
the action of the operations  
$\delta^{-1} \ep$, $co$ and  $s \in SL(2,\C)$  
generates the semidirect product structure 
$\I_2^T \times_s ( \I_2^{CPT} \times SO^+(1,3) ) \cong O(1,3)$ 
on the Minkowski space. We take the operation of $\delta^{-1} \ep$ 
and its three partners as the action of $T$ on the Weyl spinor spaces. 

It should be remarked that 
$\ep^{-1} \delta = -\delta^{-1} \ep$ and 
therefore $T^2 \cong - \1_V$ \cite{Wig 31} on the spinors, but 
$T^2 \cong \1_{\sM}$ \,on Minkowski space. 

Finally the combination $co_V \circ \ep^{-1} \delta$ is an $SU(2)$ compatible
linear isomorphism between the vector space and the anti-space. Its induced
action on the Minkowski space is given by 
\beq
	(co \ep^{-1} \delta) \bullet x \ = \ x_\mu \Lambda(P)^\mu_\nu 
		{\bf e}^\nu
\eeq
with $\Lambda(P)^\mu_\nu = 
{\scriptsize \left( \begin{array}{cc} 1 & \\ & -\1_3 \end{array} \right) }$. 
According to the remarks at the
end of sect.\ref{KO13} and for later consistency we identify this 
operation with $CP$.  

The operations dual to the $x_\mu$ are the momenta $i p^\mu = \partial^\mu$. 
Their Cartan representation lies in the dual space of the Minkowski space. 
Thus the momenta are symmetric with respect to the canonical conjugation:
\beq
	p \ \in \ i \M^T \ = \ 
		\left\{ p \in \ol{V}^T \otimes V^T | p^\times = p \right\}.
\eeq

The transformation properties of the momenta are equal to those of 
space-time for the linear operation $CP$,
\beq
	CP \bullet p \ = \ p^\mu \Lambda(P)_\mu^\nu {\bf e}_\nu 
		\ \cong \ (p^0, - \vec{p}) ,
\eeq
and different for the anti-linear operations, 
\beq
	CPT \bullet p \ = \  p , \eeq \beq
	T \bullet p \ \cong \ (p^0, - \vec{p}) .
\eeq
The anti-linear operations guarantee the positivity of the energy component
$p^0$ even in the case when time is reversed. This feature (on the level 
of the
Schr\"odinger theory) was the starting point for Wigner to define the time
reversal operation to be anti-linear \cite{Wig 31,Wig 32}. 

\section{Euclidian Conjugation and Space-Time 
Decomposition}\label{EK}
A decomposition of Minkowski space into space and time is given when there is a
distinct time-like (basis) vector.
Again it is the euclidian conjugation which provides this basis vector for
the time translations: Notice that the operation $\bar{\delta} \circ co_V$ is a
linear mapping between left- and right-handed Weyl vector spaces. 
Thus it is an
element of $\ol{V}^T \otimes V^T$. Its inverse, as an element of $V \otimes
\ol{V}$, multiplied with an $i$ can be defined as the time-like basis vector 
\beq
	{\bf e}^0 \ := \ i co_V^{-1} \bar{\delta}^{-1} .
\eeq
In the Weyl basis this vector is given by
\beq
	{\bf e}^0 \ \cong \ i \rho^0,
\eeq
The properties of a time-like basis vector of Minkowski space 
\beq
	\left({\bf e}^0 \right)^\times \ = \ - {\bf e}^0   \eeq \beq
	<{\bf e}^0 | {\bf e}^0 > \ = \ 1
\eeq
have to be proven without using any basis. This more technical part is 
done in app.\ref{AB}. The time translations are given by 
$\T = \R \cdot {\bf e}^0$ 
and space is its orthogonal complement with respect
to the Lorentz bilinear form $\Sp = \T^{\bot}$. But there is no basis
distinguished within position space. 

\section{Discrete Symmetry Operations on the Weyl Spinor Fields} \label{WS}
We show in this section how the associations of the discrete symmetries in
sect.\ref{S3} lead to
the well known operations on Weyl- and Dirac spinor fields.

The left and right handed Weyl spinors are elements of the complex Weyl
quartet: $l \in V, l^\dagger \in \ol{V}, r \in \ol{V}^T, r^\dagger \in V^T$.
The spinor fields are mappings from Minkowski space $\M$ into these 
vector spaces, e.g. $l(\cdot) \in V^{\sM}$, carrying a positive unitary
representation of the Poincar\'e group. Massive Weyl spinor fields have as 
harmonic analysis in the Wigner representation \cite{Nie}
\be
	l^A(x) &=& \int \frac{d^3q}{(2 \pi)^{3/2}} \sqrt{\frac{m}{q_0}} 
		s(\vec{q},m)^A_B \frac{e^{iqx} u^B(\vec{q}) + e^{-iqx} 
		a^{\star B}(\vec{q}) }{\sqrt{2}}  \\
	l^\dagger_{\dot{A}}(x) &=& \int \frac{d^3q}{(2 \pi)^{3/2}} 
		\sqrt{\frac{m}{q_0}} \bar{s}(\vec{q},m)^B_{\dot{A}} 
		\frac{ e^{-iqx} u^\star_B(\vec{q}) + e^{iqx} a_B(\vec{q}) 
		}{\sqrt{2}}  \\
	r^{\dot{A}}(x) &=& \int \frac{d^3q}{(2 \pi)^{3/2}} \sqrt{\frac{m}{q_0}} 
		\check{\bar{s}}(\vec{q},m)^{\dot{A}}_B 
		\frac{ e^{iqx} u^B(\vec{q}) - e^{-iqx} a^{\star B}(\vec{q}) 
		}{\sqrt{2}}  \\
	r^\dagger_A(x) &=& \int \frac{d^3q}{(2 \pi)^{3/2}} \sqrt{\frac{m}{q_0}} 
		\check{s}(\vec{q},m)^B_A \frac{e^{-iqx} u^\star_B(\vec{q}) - 
		e^{iqx} a_B(\vec{q}) }{\sqrt{2}} . 
\ee
Here $s(\vec{q},m) = D(s)$ is the matrix representation for a representative $s$ 
of a boost coset $SL(2,\C)/SU(2)$ 
parametrized by the momenta $q$ of the mass $q^2 = m^2, m > 0$ \cite{Sal}:
\be
	s^A_B(\vec{q},m) & = & \sqrt{\frac{q_0 + m}{2m}} \ \left[ 
		{\1_2} + \frac{\vec{\sigma} \cdot \vec{q}}{q_{\rm 0}+m} 
		\right] \nonumber  \\
		& = &
		\frac{1}{\sqrt{2m(q_0 + m)}} \left( \begin{array}{cc}
			q_0 + m + q_3 & q_1 - iq_2 \\
			q_1 + i q_2 & q_0 + m - q_3 
		\end{array}  \right) .
\ee
$u(\vec{q}), u^\star(\vec{q})$ and $a(\vec{q}), a^\star(\vec{q})$ are the 
creation and annihilation operators of particles and antiparticles, resp. 
According to Wigner they carry only a finite dimensional positive unitary
representation of the little group \cite{Wig 39}, which for massive 
spinor fields 
is $SU(2)$. Therefore these operators map into the complex
representation quartet\footnote{The star $\star$ denotes the
euclidian conjugation. The assignment of the creation and the
annihilation operators is reversed compared to the standard notation.} of
$SU(2)$, 
$u \in V, u^\star \in V^T, a \in \ol{V}, a^\star \in \ol{V}^T$. As
mappings they have the discrete transformation properties
induced by the discrete transformations of the momenta and of the Weyl spinors.
For the corresponding basis this is 
\be
	CPT & : & co \bullet u^A(\vec{p}) \ = \ \delta^{AB} a_B(\vec{p}) \\
	T & : & (\delta^{-1} \ep) \bullet u^A(\vec{p}) \ = \ 
		\ep^{AB} \delta_{BC} u^C(- \vec{p}) \\
	CP & : & (co \ep^{-1} \delta) \bullet u^A(\vec{p}) \ = \ 
		\delta^{AB} \ep_{BC} \delta^{CD} a_D(- \vec{p}) ,
\ee
with $\ep^{AB} \ep_{BC} = \delta^A_C$. 
Calculating the action of the discrete symmetry operations on the 
representations of the boosts
\be
	co \bullet s(\vec{q},m) &=& \bar{s}(\vec{q},m)  \\
	(\delta^{-1} \ep) \bullet s(\vec{q},m) &=& s(-\vec{q},m)  \\
	(co \ep^{-1} \delta) \bullet s(\vec{q},m) &=& \bar{s}(-\vec{q},m)  
\ee
one obtains the action of the discrete symmetries on quantized spinor 
fields (compare, e.g., \cite{SW}):  
\be
	CPT &:& \left\{ \begin{array}{ccc}
		l^A(x)^\times & := & co \bullet l^A(x) \ = \ \delta^{A
		\dot{B}} l^\dagger_{\dot{B}}(-x)  \\
		r^{\dot{A}}(x)^{\times} & := & co \bullet r^{\dot{A}}(-x) \ = \ 
		- \delta^{\dot{A}B} r^\dagger_B(-x)
	\end{array} \right. \\
	T &:& \left\{ \begin{array}{ccc}
		(\delta^{-1} \ep) \bullet l^A(x) & = & \ep^{AB} \delta_{BC} 
			l^C(-t,\vec{x})  \\
		(\delta^{-1} \ep) \bullet r^{\dot{A}}(x) & = & 
			\ep^{\dot{A}\dot{B}} \delta_{\dot{B}\dot{C}}
			r^{\dot{C}}(-t,\vec{x}) \end{array} \right. \\
	CP &:& \left\{ \begin{array}{ccc}
		(co \ep^{-1} \delta) \bullet l^A(x) & = & - \delta^{AB} 
			\ep_{BC} \delta^{\dot{C}\dot{D}}
			l^{\dagger}_{\dot{D}}(t,-\vec{x})  \\
		(co \ep^{-1} \delta) \bullet r^{\dot{A}}(x) & = & 
			 \delta^{\dot{A}\dot{B}} \ep_{\dot{B}\dot{C}} 
			 \delta^{\dot{C}D}  r_D^\dagger(t,-\vec{x}) ,
		\end{array} \right.  
\ee
Again these are induced actions. For example the $CPT$ operation on left-handed 
Weyl spinor fields is given by the commutative diagram \\
\linebreak
\setlength{\unitlength}{0.6cm}
\begin{picture}(10,6.5)(-4,0)
        \put(0.5,0.5){\makebox{$V$}}
        \put(0.5,5.0){\makebox{$\M$}}
        \put(5.5,5.0){\makebox{$\M$}}
        \put(5.5,0.5){\makebox{$\ol{V}$}}
        \put(1.0,5.1){\vector(1,0){4}}
        \put(1.0,0.8){\vector(1,0){4}}
        \put(0.6,4.5){\vector(0,-1){3}}
        \put(5.6,4.5){\vector(0,-1){3}}
        \put(2.5,5.4){\makebox{$co \bullet$}}
        \put(2.5,0.2){\makebox{$co_V$}}
        \put(-0.3,3.0){\makebox{$l(\cdot)$}}
        \put(6.1,3.0){\makebox{$l(\cdot)_{CPT}$}}
\end{picture}
\linebreak

In the Weyl representation of the Dirac field\footnote{Because $l^\dagger$ is 
dual to $r$ and $r^\dagger$ is dual to $l$, the dual to $\psi$ is the Dirac 
adjoint $\ol{\psi}$.}, 
\[
	\psi(x) \ =\  l(x) \oplus r(x) \ \cong \  
		\left( \begin{array}{c} l(x) \\ r(x) \end{array} \right) , 
\] 
\[
 	\psi^\dagger(x) \ = \ l^\dagger(x) \oplus r^\dagger(x) \ \cong \ 
 		\left( \begin{array}{cc} l^\dagger(x) ,& r^\dagger(x) 
 		\end{array} \right)  ,
\]
the action of the linear and anti-linear discrete operations can be expressed 
with the Dirac matrices given naturally in the chiral or Weyl representation, 
\be
	CPT \bullet \psi(x) & \cong & \psi^\dagger(-x) \gamma^5 \\
	T \bullet \psi(x) & \cong & \gamma^1 \gamma^3 \psi(-t,\vec{x})  \\
	CP \bullet \psi(x) & \cong & \psi^\dagger(t, -\vec{x}) 
		\gamma^0 i \gamma^2. 
					\label{CPDF}
\ee
So we recover the discrete symmetry operations of quantum field theory 
on the Dirac spinor fields \cite{Lud, BD2}. 

\section{Discrete Symmetry Operations and Inner Symmetries}
The spinor fields of the Standard Model have nontrivial inner (gauge) symmetries.
They are mappings onto a tensor product space $V \otimes U$ of a 
representation space $V$ for the Lorentz symmetry and a 
representation space $U$ for the inner symmetry. To extend the discrete
symmetries on this tensor product space we have to continue the operations
$co$, $\delta$ and $\ep$ on the inner symmetry space compatible with the inner
symmetry. This is trivial for the canonical conjugation $co$, since its 
canonical construction does not depend on the symmetry structure. The euclidian
conjugation $\delta$ can be continued on every inner symmetry space, because 
the inner
symmetries are positive unitary groups, the invariance 
groups of the euclidian conjugation (eq.(\ref{Un})). 
In general, however, this is not possible for the $SU(2)$ and $SL(2,\C)$
invariant 
volume form $\ep$, because for more than two complex dimensions $\ep$ is 
not bilinear. 

To be more explicit, let us focus on some fields used in the Standard Model.
The left-handed leptons carry a left-handed Weyl representation of the Lorentz
group and the fundamental representation of $SU(2)$ weak isospin. 
They are elements of a tensor product space 
${\fett{l}} \in V_l = V \otimes U_2 \cong \C^2 \otimes \C^2$. 
The canonical conjugation
\beq
	co_{V_l} : \ V_l \longrightarrow \ol{V}_l = \ol{V} \otimes \ol{U}_2
\eeq
defines the representations of the Lorentz group and the weak isospin on the
anti-space. Thus $CPT$ is per definition compatible with the symmetry
structure. For the time reversal we need an $SU(2)_{\mbox{spin}} \times
SU(2)_{\mbox{isospin}}$ compatible anti-linear automorphism. This is possible
if we use the weak isospin volume form $\ep_{U_2}$, with the action
\be
	T \bullet \fett{l} & = & \left( \delta^{-1}_{V_l} \circ 
		\ep_{V} \otimes \ep_{U_2} \right)  (\fett{l})  \\  
	CP \bullet \fett{l} & = & \left( co_{V_l} \circ 
		(\ep_{V} \otimes \ep_{U_2})^{-1} \circ \delta_{V_l} 
			\right) (\fett{l}) .
			\label{GP}
\ee
These operations include an interchange between the two weak isospin 
components (after spontaneous symmetry breaking they can be identified for
example with the
neutrino and the electron). This introduction of $\ep_{U_2}$ is equivalent 
to the introduction of $G$-parity for the strong isospin
\cite{Nis 51,Mic 53,Lee 56}. Hence eq.(\ref{GP}) defines  
a weak isospin $GP$ operation. 

The situation changes drastically if we include symmetries $SU(n)$ with $n \geq
3$, like colour-$SU(3)$. As the simplest example we use 
the right-handed quarks. They carry a right-handed Weyl representation of the
Lorentz group and a fundamental triplet representation of $SU(3)_{\mbox{colour}}$. 
They are
elements of a tensor product space ${\bf q} \in V_q = \ol{V}^T \otimes U_3 
\cong \C^2 \otimes \C^3$. The canonical conjugation again defines $\ol{V}_q$
and its representation structure, thus $CPT$ is defined. 
Since there is no $SU(3)$ invariant bilinear form on $U_3$ with which one could
extend the operations $T$ and $CP$ on the inner symmetry space, 
$T$ and $CP$ cannot be defined compatibly with colour-$SU(3)$. 

To be even more explicit, take the fundamental matrix representations of the 
gauge groups $SU(2)$ and $SU(3)$ given by the 
Pauli- and Gell-Mann-matrices, resp. 
\be
	D(u_2).{\bf l} & = & e^{\frac{i}{2} \alpha_j \tau_j} {\bf l} , 
		\ \ \ j = 1, .. , 3 \  , \  \alpha_j \in \R \\\
	D(u_3).{\bf q} & = & e^{\frac{i}{2} \beta_a \lambda_a} {\bf q} , 
		\ \ \ a = 1, .. , 8 \ , \ \beta_a \in \R  .  
\ee
Then the representation of $SU(2)$ and $SU(3)$ on the anti-spaces (in this case
equivalent to the dual representation) is given by
\be
	\bar{D}(u_2).{\bf l}^\dagger & = & e^{- \frac{i}{2} \alpha_j \bar{\tau}_j} 
		{\bf l}^\dagger ,  \\
	\bar{D}(u_3).{\bf q}^\dagger & = & e^{- \frac{i}{2} \beta_a 
		\bar{\lambda}_a} {\bf q}^\dagger , 
\ee
with e.g. $\bar{\tau}_j$ denotes the conjugation of the entries in the matrix
without transposition. 
The action of $CP$ is a linear operation between these two representation
spaces. A compatible $CP$ operation has to fullfill 
\beq
	CP \bullet \left( D(u).\psi \right) \ = \ \bar{D}(u).\psi_{CP} 
\eeq
For the $SU(2)$ representation this is possible with the matrix representation
of $\ep_{U_2}$ usually denoted by\footnote{This sloppy notation seems to
distinguish a basis. Its use is justified only by the identical matrix
of the dual isomorphism $\ep_{U_2}$ and the endomorphism 
$i \tau_2$.}
$i \tau_2$ (for simplicity the Lorentz structure is omitted)
\beq
	{\bf l}_{CP} \ = \ i \tau_2 {\bf l}^\dagger  \eeq \beq
	\Rightarrow \ e^{- \frac{i}{2} \alpha_j \bar{\tau}_j} i \tau_2 
			{\bf l}^\dagger 
		\ = \ i \tau_2 \left( e^{\frac{i}{2} \alpha_j \tau_j} 
			{\bf l}^\dagger \right) ,
\eeq
but there is no linear operation for the $SU(3)$ representation corresponding
to $\tau_j\  =\  i \tau_2\ \bar{\tau}_j \ i \tau_2$.

\section{Treatment of $CP$ in the Standard Model}
The interaction Lagrangian of a gauge theory is believed to be invariant
under $CP$ transformation. The $CP$ violating part in the Standard Model is 
provided by the mixing of
the three families via the Cabibbo-Kobayashi-Maskawa matrix \cite{Kob 73}.
Whereas the invariance of the interaction Lagrangian seems to
contradict our analysis, we can show that it agrees with the KM-theory. 
For the standard treatment of $CP$ we follow \cite{BD2, Nach, Jar}. 

From the action of $CP$ on Dirac spinor fields, eq.(\ref{CPDF}), one can
calculate 
the action on all bilinear products, especially for the $U(1)$-current $j^\mu =
\frac{1}{2} [\bar{\psi},\gamma^\mu \psi ]$, 
\beq
	CP \bullet j^\mu(x) \ = \ CP \bullet \frac{1}{2} 
			[\bar{\psi}(x),\gamma^\mu \psi(x) ]
		\ = \ - \frac{1}{2} 
			[\bar{\psi}(t,\vec{x}),\gamma_\mu \psi(t,\vec{x}) ] 
		\ = \ - j_\mu(t,\vec{x})  .
\eeq
Including inner degrees of freedom the currents $j^\mu_{ij} =
\frac{1}{2} [\bar{\psi}_i,\gamma^\mu \psi_j ]$, 
with $\psi_i$ is a (basis) vector in the 
inner symmetry space and $i,j$ are the inner indices, transform according to 
\be
	CP \bullet j^\mu_{ij}(x) & = & 
		CP \bullet \frac{1}{2} 
			[\bar{\psi}_i(x),\gamma^\mu \psi_j(x) ]  \nonumber
			\\
		& = & - \frac{1}{2} 
			[\bar{\psi}_j(t,-\vec{x}),\gamma_\mu \psi_i(t,-\vec{x}) ]
			\nonumber \\
		& = & - j_{\mu ji}(t,-\vec{x})  .
\ee
This current couples to the Lie-algebra valued gauge fields 
\[
	G^\mu_{ij}(x) = G^{a \mu}(x) {\cal D}(l_a)_{ij}  .
\]
with $G^{a \mu}(x)$ the gauge field and ${\cal D}(l_a)$ a matrix 
representation of the gauge Lie-algebra. Choosing the transformation properties
of the gauge fields in such a way that 
\be
	CP \bullet G^\mu_{ij}(x) & = & - G_{\mu ji}(t,-\vec{x}) \nonumber \\
		& = & - G^{a}_{\mu}(t,-\vec{x}) {\cal D}(l_a)_{ji} 
			\nonumber \\
		& =: & G'^{a}_{\mu}(t,-\vec{x}) {\cal D}(l_a)_{ij} ,
\ee
e.g. for the representation of $su(2)$ with the Pauli matrices 
\beq
	CP \ : \ \left( G^{1 \mu}, G^{2 \mu}, G^{3 \mu} \right)(x) \ 
		 \longmapsto \
		- \left( G^{1}_{\mu}, -G^{2}_{\mu}, G^{3}_{\mu} \right)
			(t,-\vec{x}) , \label{su2}
\eeq
and for the representation of $su(3)$ with the Gell-Mann matrices 
\[
	CP \ : \ \left( G^{1 \mu}, G^{2 \mu}, G^{3 \mu}, G^{4 \mu}, 
			G^{5 \mu}, G^{6 \mu}, G^{7 \mu}, 
			G^{8 \mu} \right)(x) \] \beq \longmapsto \ 
		- \left( G^{1}_{\mu}, -G^{2}_{\mu}, G^{3}_{\mu}, G^{4}_{\mu}, 
			-G^{5}_{\mu}, G^{6}_{\mu}, -G^{7}_{\mu}, G^{8}_{\mu}
			\right)(t,-\vec{x}) .
				\label{su3} 
\eeq
This leaves the Lagrange density formally invariant. 

There are two points of criticism.

First, the $CP$ transformations for the gauge bosons are basis-dependent: 
the transformations (\ref{su2}) and (\ref{su3}) are given only with the
representation of $su(2)$ and $su(3)$ by the Pauli- or the Gell-Mann matrices,
resp.\footnote{A basis-independent operation for the Lie algebra representation
of $su(2)$ for $G^\mu \mapsto {G^\mu}^T$ is given only by the volume
form $\ep_{U_2}$ 
\[
	\ep_{U_2} \circ {\cal D}(l) \circ \ep_{U_2}^{-1} \ = \
		\check{\cal D}(l) \ = \ - {\cal D}(l)^T  .
\]
This is the action of the $GP$-operation we introduced in (\ref{GP}) on 
the endomorphisms.} Another representation would yield another
transformations of the gauge fields. 

Whereas in the elektroweak sector via the Higgs field there exists a
distinguished basis 
(asymptotically we know the difference between electron and neutrino),
this is believed not to be the case for the quarks. The assumption of a 
distinguished basis in the
colour-space contradicts the concept of an unbroken $SU(3)$ gauge theory.

Secondly, an operation on the Lie-algebra representation without referring to its
action on the vector space of the representation is, at least from an algebraic
point of view, unsatisfactory. Vice versa, the action on the endomorphisms are 
uniquely determined by the action on the representation space via the inducing
construction. 

The problem of $CP$-violation is treated in the Standard Model in 
terms of the three families and
their mixing by the Cabibbo-Kobayashi-Maskawa matrix. The CKM
matrix is an unitary basis transformation of a three dimensional family
space $F \cong \C^3$. Neglecting the inner symmetries the down-type quarks for
example are elements of a tensor product space $V_d = V \otimes F \cong \C^2
\otimes \C^3$. The mathematical structure is similar to the case of the 
right-handed quarks
with the only difference, that in $F$ there are fixed bases - one for to the 
mass eigenstates and one for to the weak interaction (current
eigenstates) -  correlated by the CKM matrix.  
Just as $CP$ is not compatible with
$SU(3)_{\mbox{colour}}$, it is not compatible with the $U(3)$ CKM matrix,
either. 
Because $SO(3)$ has an invariant dual isomorphism $CP$ would not be
violated if the CKM matrix would be orthonormal. In this sense our
analysis coincides with the Kobayashi-Maskawa theory. 

\section{Conclusion}
Without a separation of space and time there is a $CPT$-action but no $CP$ and 
$T$-actions.

Therefore in a theory without massive asymptotic states like pure QCD there is 
no need in defining operations for $CP$ and $T$. Hence, the impossibility
of defining a $CP$ operation compatible with $SU(3)$ might be of no
phenomenological consequence. For the colourless asymptotic 
particles of the strong interaction, the massive hadrons, $CP$ is well defined. 
It would be of
interest whether the impossibility of defining an $SU(3)$-compatible $CP$ 
operation leads, via the Higgs mechanism or the confinement, to $CP$ violation 
given
by the family-mixing of the KM-theory. A hint may be the similarity in the 
mathematical structure of the three colours and the three families. 

The (mathematical) difference of the $SU(3)$ gauge theory
and the KM-theory with respect to $CP$ is the
nonexistence of a basis in the colour-space whereas there are two distinct
bases in the family-space. This situation can be visualized easier in the 
elektroweak sector. Here in the pure $U(1) \times SU(2)$ gauge theory there is 
no basis given in the representation space, i.e.\,there is no difference 
between electron- and neutrino fields.
Spontaneous symmetry breaking distinguishes bases for
the Higgs field in the inner symmetry space. With the breaking
of $SU(2)_L \times U(1)_Y$ to $U(1)_Q$ comes along the concept of mass and
therewith the difference of electron- and neutrino particles. 
Hence the basis depending operation of (\ref{su2})
becomes possible after spontaneous symmetry breaking, 
i.e. for the asymptotic states. 

\section*{Acknowledgment}
The authors would like to thank H.\,Tann and B.\,Fauser for discussion and
critical reading of preliminary versions of the manuscript.

\begin{appendix}
\section{The Lorentz Bilinear Form of the Cartan Representation of Minkowski
space}  \label{AA}
Every dual isomorphism is equivalent to a bilinear form. The action of the
spinor metric $\ep$ on the Cartan representation of Minkowski space 
\beq
	\ep \bullet x \ = \ \bar{\ep} x^T \ep \ \in \ \M^T
\eeq
is again a dual isomorphism. We show that this is equivalent to the Lorentz
bilinear form by calculating the action of this dual vector on the Minkowski
vectors:
\be
	< x | y >_\ep &=& (\ep \bullet x) (y) \nonumber \\
		& = & \frac{1}{2} \mbox{tr} \bar{\ep} x^T \ep y 
			\label{LBF}
\ee
Especially in the Weyl basis this gives the familiar form of the Lorentz
bilinear form
\be
	< x | y >_\ep &\cong& \frac{1}{2} \tr \left[ 
		\left( \begin{array}{cc}
			0 & 1 \\
			-1 & 0 
			\end{array} \right)
		\ i \left( \begin{array}{cc}
			x_0 + x_3 & x_1 - i x_2 \\
			x_1 + i x_2 & x_0 - x_3 
			\end{array} \right)^T \right.  \times
		\nonumber \\ && \ \ \ \ \left. \times
		\left( \begin{array}{cc}
			0 & 1 \\
			-1 & 0 
			\end{array} \right)
		\ i \left( \begin{array}{cc}
			y_0 + y_3 & y_1 - i y_2 \\
			y_1 + i y_2 & y_0 - y_3 
			\end{array} \right)
		\right] \nonumber \\
	&=& x_\mu \eta^{\mu\nu} y_\nu,  \nonumber
\ee
with $\eta = 
{\scriptsize \left( \begin{array}{cc} 1 & \\ & -\1_3 \end{array} \right) } $.

\section{The Time-like Basis Vector ${\bf e}^0$} \label{AB}
In sect.\ref{EK} we defined as time-like basis vector 
${\bf e}^0 = i co_V^{-1} \bar{\delta}^{-1}$. Here we prove the properties 
\be
	1. && \left({\bf e}^0 \right)^\times  \ = \ -e^0 \\
	2. && <{\bf e}^0 | {\bf e}^0 > \ = \ 1  .
\ee
by refering only to the properties of $co_V, \delta$
and $\ep$. 

The first equation writes  
\be
	\left({\bf e}^0 \right)^\times & = & \left( i co_V^{-1} \bar{\delta}^{-1} \right)^\times
			\nonumber \\
		& = & co_V^{-1} \left( i co_V^{-1} \bar{\delta}^{-1} \right)^T
			co_{V^T}^{-1} \nonumber \\
		& = & -i co_V^{-1} \bar{\delta}^{-1} co_{V^T} co_{V^T}^{-1}
			\ = \ - {\bf e}^0.  \nonumber
\ee
	
For the second equation we have to compute the square of $e^0$ with respect to
the Lorentz bilinear form in its basis independent definition of eq.(\ref{LBF}):
\be
	 <{\bf e}^0 | {\bf e}^0 > & = & \frac{1}{2} \mbox{tr}\ \bar{\ep}\ {{\bf e}^0}^T\ 
			\ep\ {\bf e}^0  \nonumber \\
		& = & \frac{1}{2} \mbox{tr}\ \bar{\ep} \left( i co_V^{-1} 
			\bar{\delta}^{-1} \right)^T  \ep\ i co_V^{-1}\ 
			\bar{\delta}^{-1} \nonumber \\
		& = & - \frac{1}{2} \mbox{tr}\ \bar{\ep}\ \bar{\delta}^{-1}\ 
			co_{V^T}\ \ep\ co_V^{-1}\ \bar{\delta}^{-1} \nonumber \\
		& = & - \frac{1}{2} \mbox{tr}\ \bar{\ep}\ \bar{\delta}^{-1}\ 
			\bar{\ep}\ \bar{\delta}^{-1} \nonumber \\
		& = & - \frac{1}{2} \mbox{tr} \left(- \1_{\ol{V}^T} \right)  
			\ = \ 1 .\nonumber
\ee
\end{appendix}

\end{document}